\documentstyle[11pt,newpasp,twoside,epsf]{article}
\def\edcomment#1{\iffalse\marginpar{\raggedright\sl#1\/}\else\relax\fi}
\reversemarginpar

\begin{document}
\title{The National Virtual Observatory}
\author{Robert J. Brunner, S. George Djorgovski}
\affil{Palomar Observatory, California Institute of Technology, Pasadena, CA 91125}
\author{Thomas A. Prince}
\affil{Space Radiation Laboratory, California Institute of Technology, Pasadena, CA 91125}
\author{Alex S. Szalay}
\affil{Department of Physics and Astronomy, Johns Hopkins University, Baltimore, MD 21218}

% Add as many "\author" and "\affil" fields as necessary.

\begin{abstract}
As a scientific discipline, Astronomy is rather unique. We only have
one laboratory, the Universe, and we cannot, of course, change the
initial conditions and study the resulting effects. On top of this,
acquiring Astronomical data has historically been a very
labor-intensive effort. As a result, data has traditionally been
preserved for posterity. With recent technological advances, however,
the rate at which we acquire new data has grown exponentially, which
has generated a Data Tsunami, whose wave train threatens to overwhelm
the field. In this conference proceedings, we present and define the
concept of virtual observatories, which we feel is the only logical
answer to this dilemma.
\end{abstract}

\section{The Problem}

A major paradigm shift is now taking place in astronomy and space
science. Astronomy has suddenly become an immensely data-rich field,
with numerous digital sky surveys across a range of wavelengths, with
many Terabytes of pixels and with billions of detected sources, often
with tens of measured parameters for each object.  This is a great
change from the past, when often a single object or small samples of
objects were used in individual studies.  Instead, we can now map the
universe systematically, and in a panchromatic manner.  This will
enable quantitatively and qualitatively new science, from statistical
studies of our Galaxy and the large-scale structure in the universe,
to the discoveries of rare, unusual, or even completely new types of
astronomical objects and phenomena. This new digital sky, data-mining
astronomy will also enable and empower scientists and students
anywhere, without an access to large telescopes, to do first-rate
science.  This can only invigorate the field, as it opens the access
to unprecedented amounts of data to a fresh pool of talent.

The resulting wealth of data from the wide range of surveys, satellite
missions, and ground-based observatories provides enormous new
scientific opportunities. Archiving, distributing and exploring all of
this data, however, is an extremely challenging, yet vital,
problem. This fact was recognized by the National Academy of Science
Astronomy and Astrophysics Survey Committee, which, in its new decadal
survey (Astronomy and Astrophysics in the New
Millennium\footnote{http://www.nap.edu/books/0309070317/html}),
recommends, as a first priority, the establishment of a National
Virtual Observatory. We consider this an extremely important edict,
and encourage the rest of the astronomical community to work towards
the successful implementation of not just a single national facility,
but a global network of interconnected virtual observatories. After
all, the universe does not know of our artificially imposed
geographical boundaries, everyone suffers from the data overload.

At a fundamental level, several complimentary effects drive this data
tsunami. Primary among these is the incredible changes that have
arisen from the application of new technology. Moore's law, a
well-known observation, roughly states that silicon-based device
capacity doubles every 18 months.  The most noticeable effect from
this observation is the incredible advances in computational speed and
power, something that is most definitely not limited to the field of
astronomy.  As a result, we have storage systems that allow us to
archive our data, and the growth of the Internet and the World Wide
Web simplifies the process of accessing highly distributed data. On
the other hand, Moore's law is also driving the growth in detector
technology; Giga-pixel arrays are on the horizon and even more
powerful devices are in the pipeline, resulting in ever increasing
data streams. This data tsunami is not a one-time event, the driving
mechanisms continue to operate and increase in amplitude.

Another important driver results from the globalization of
astronomy. Over the last decade there has been a tremendous growth in
both the number and power of telescopes being built around the
world. At the same time, a range of multi-wavelength space missions is
pushing the limits of our understanding, and providing new views on
our cosmos. In the end, we will not only have the electromagnetic
spectrum opened before us, but also the gravitational, astroparticle
and time domains. Not to be left behind, computer simulations are
rivaling the largest observational datasets in both size and
complexity, providing new approaches to understanding the universe.

From the previous discussion, the current state of astronomy would
seem to be glorious. The true situation, however, is hampered by
several limiting forces, most notably the human factor. Funding for
the scientific exploration of the wealth of data that is inundating
our field is inadequate to the task. Furthermore, astronomers are not
trained to handle such large datasets, and, as a result, the data that
is so expensively obtained is not completely explored. Finally, since
general data archiving and distribution facilities, especially of
homogeneous, large area surveys are not commonly available,
observations are often repeated, either due to ignorance or by
necessity, as a result of geographical, cultural, or proprietary
forces.

\section{The Solution}

At some level, everyone is familiar with the awe-inspiring pageantry
of the night sky, and most people are familiar with the basics of a
telescope. For the professional astronomer, obtaining astronomical
data from the ground is generally performed at observatories that are
often located on remote mountain peaks. These observatories support
the discovery process by providing access to telescopes and
instruments that are suited for different types of observations. At a
fundamental level, the telescope's sole function is to gather photons
and feed them to the appropriate instrument, which is designed to
analyze the incoming photons and to, hopefully, extract the desired
signal. There are a large number of observatories, distributed
worldwide, which share information on how to build and improve
telescopes and instrumentation, and, occasionally, the actual physical
hardware.

This description provides a useful framework for building a mental
picture of how a network of Virtual Observatories would
operate. Rather than starting with the physical universe, we start
with the digital universe, consisting of the vast quantities of
archived data, from both ground- and space-based observatories. In
order to explore this new paradigm, we need virtual observatories,
which serve as portals into the digital realm. Their primary function
would be to provide access to virtual telescopes that serve to gather,
not photons, but the bits of information that are relevant to a given
query from the vast sea of information. Virtual instruments will be
used to process the resulting data stream, extracting out the desired
result. Due to their inherent ``soft'' nature, these virtual telescopes
and instruments can be easily reconfigured, expanded, and ultimately
improved in order to tackle the next data discovery challenge.

With this mental paradigm in place, we can now more clearly define a
``Virtual Observatory''. First, a virtual observatory would be global in
nature, distributed in scope across institutions, agencies, and
countries. In addition, a virtual observatory being based on the
Internet would be continuously available to everyone, both in the
astronomical community, as well as the general public, wherever they
might be located. Unlike traditional observatories, the infrastructure
for a virtual observatory can not be physically located in a single
location, following the ``click-and-order'' not ``brick-and-mortar''
philosophy which drives the net-economy. This is a necessary
consequence of the fact that virtual observatories would support
astronomical observations via remote access to digital representations
of the sky. Finally, virtual observatories must provide general
support for astronomical explorations of large areas of the sky at
multiple wavelengths, and enable discovery via new computational and
statistical tools.

The architecture for a virtual observatory, would, therefore, follow a
grid-based pattern, where different data providers, which are broadly
distributed in both geography and resources, are interconnected to
both the user community, and specialized compute servers. This
hierarchical approach allows existing services that have developed
over time to support individual satellite missions, observatories, or
surveys to be easily integrated, while simultaneously allowing for
future growth.

Other scientific disciplines are experiencing similar deluges of data,
and we need to work together, maximizing our joint resources to
solve our common problems. We also need to develop and promote
collaborations with computer scientists and statisticians, who have
the necessary skills to develop the next generation data handling
tools and techniques that will provide the infrastructure and core
functionality of a world of virtual observatories.

\section{The Result}

The end result of this evolutionary process is that astronomy has now
become an information science. With the new capabilities provided by
virtual observatories, exciting new scientific avenues can be
explored. With the federation of the enormous quantity of
multi-wavelength observations, we can discover and study in more
detail rare and exotic objects such as high redshift quasars, obscured
quasars, extreme broad absorption line quasars, T and L sub-dwarfs,
brown dwarfs, and ultra-cool white dwarfs. The vast quantity of
homogeneous survey data will promote the development of new, higher
precision studies of the structure of our galaxy and universe, as
systematic errors, as opposed to statistical errors, become the
limiting factor.

Another new possibility would be to harness the available compute
power to work directly in the image domain, to perform
multi-wavelength source detection and extraction. We also will be able
to correlate diffuse emission across wavelengths to look for
structures that more traditional techniques overlook. Finally, perhaps
the most exciting scientific possibility is the opening of the time
domain, which would not only allow the study of known classes of
objects such as gravitational micro-lensing, supernovae, and
near-earth asteroids, but also allow for the exploration of a
hither-to barely explored area of parameter space.

While the benefits to the scientific community are impressive on their
own, virtual observatories can revolutionize how we as a community
interact with the public. Astronomy is rather unique in its wide
appeal; by providing broad public access to panchromatic images and
simulations of the ever-changing sky, virtual observatories can convey
the importance of astronomy, and science in general, to the public and
advance overall science literacy. Instructors at all grade levels
could interface their curriculum with a virtual observatory, allowing
students to participate in a ``hands-on'' manner in the acquisition and
analysis of data. The possibilities truly are limitless.

It is quite obvious that handling and exploring these vast new data
volumes, and actually making real scientific discoveries poses a
considerable technical challenge.  The traditional astronomical data
analysis methods are inadequate to cope with this sudden increase in
the data volume (by several orders of magnitude).  Fortunately, the
existing computing technology can provide such tools, both in hardware
and software, but their large-scale implementation in astronomy has
not yet happened.

As a result, we expect that the major challenge facing the
implementation of virtual observatories is not technological in
origin, but sociological.  This is a natural result of the enormous
consequences for the astronomical community posed by virtual
observatories, which provide for an entirely new way of doing
astronomy. All areas of astronomy will be affected, as people
everywhere will suddenly have access to huge quantities of data and
the tools necessary to explore it --- a process that we call the
democratization of astronomy.

We expect that the astronomical community will be very polarized by
this concept; those who embrace the concepts we are promoting, and
those who either fear the forthcoming change, or do no think it will
ever work. As a result, we feel that while virtual observatories will
be enabled by technology, their development must be driven by science,
in order to win broad community support. Together with the enormous
new opportunities a virtual observatory presents, science drivers will
serve to train the community, and in particular, the next generation
of astronomers, to capitalize on the revolution which is sweeping our
community.

\section{Summary}

Astronomers have a long history of incorporating new technologies into
their research. Eventually, Virtual Observatory capabilities will
become available to the community, and, just as importantly, the
general public. A very successful conference was recently held on the
campus of the California Institute of Technology, which explored the
impact Virtual Observatory technologies would have on our field. We
encourage anyone who is interested to explore the proceedings from
that conference (Virtual Observatories of the Future) or the
conference's
web-site\footnote{http://www.astro.caltech.edu/nvoconf}. In addition,
an NVO white paper (NVO White Paper 2001) was developed by interested
personnel within the US astronomical community, which provides more
details than could be presented in this conference proceedings. All
correspondence should be directed to the the lead author
(rb@astro.caltech.edu).

\acknowledgements

We are grateful to all of our collaborators from around the world who
share our vision. In addition, we wish to thank NASA and NSF for their
encouragement in difficult times, and both SUN Microsystems and
Microsoft Research for their support. RJB would like to explicitly
acknowledge financial support from NASA grants NAG5-10885 and NAG5-9482.


\begin{references}

\reference{Astronomy and Astrophysics Survey Committee, 
{\em Astronomy and Astrophysics in the New Millennium}, 
the National Academy Press, Washington, DC.}

\reference{NVO White Paper, in {\em Virtual Observatories of the Future}, 
eds. R.J. Brunner, S.G. Djorgovski, and A.S. Szalay, ASP conference series, 225, 353.}

\reference{{\em Virtual Observatories of the Future}, 
eds. R.J. Brunner, S.G. Djorgovski, and A.S. Szalay, ASP conference series, 225.}

\end{references}
\end{document}